\documentclass[10pt,letterpaper]{article}

\usepackage[letterpaper,hmargin=1.25in,vmargin=1.25in]{geometry}
\usepackage{natbib}
\usepackage{amssymb, latexsym, amsmath, amsthm, mathrsfs}
\usepackage{amsbsy}
\usepackage{graphicx}
\usepackage[usenames]{color}
\usepackage{ifthen}
\usepackage{soul}
\usepackage{setspace}
\usepackage{rotating}
\usepackage{placeins}
\theoremstyle{plain}
\newtheorem{theorem}{Theorem}

\newtheorem{lemma}{Lemma}

\theoremstyle{definition}

\theoremstyle{remark}
\newtheorem{remark}{Remark}                                                                                                                                                                                                                                                                                                                                                                                                                                                                                   

\renewcommand{\phi}{\varphi}

\begin{document}

\title{ A Sandwich Likelihood Correction for Bayesian Quantile Regression based on the Misspecified Asymmetric Laplace Density}
\author{  {\sc Karthik Sriram}\footnote{Karthik Sriram is Assistant Professor, Production and Quantitative Methods Area, Indian Institute of Management Ahmedabad, India ({\tt karthiks@iimahd.ernet.in})}  }


\date{}
\maketitle
\thispagestyle{empty}

\begin{center}Last revised on \today\end{center}
\doublespacing

\begin{abstract}
A sandwich likelihood correction is proposed to remedy an inferential limitation of the Bayesian quantile regression approach based on the misspecified asymmetric Laplace density, by leveraging the benefits of the approach. Supporting theoretical results and  simulations are presented.
\smallskip

\end{abstract}

\section{Introduction}
Quantile Regression is used to model conditional quantiles of the independently distributed responses $\{Y_i\}_{i=1}^n$ given $p-$dimensional covariate vectors $\{{\bf X}_i\}_{i=1}^n$. The frequentist approach to modeling the $\tau^{th}$quantile $(0<\tau<1)$ as proposed by \cite{koenkerbasset1978} involves solving the problem:
\begin{eqnarray}
\boldsymbol{\beta}^{M}_n= arg\min_{\boldsymbol{\beta}} \sum_{i=1}^{n}\rho_{\tau}(Y_{i}-{\bf X}_{i}^{T}\boldsymbol{\beta}) \label{qrfreq},
\end{eqnarray}
where $\rho_{\tau}(u)= u(\tau - I_{(u \le 0)})$ with $I_{(\cdot)}$ being the indicator function.  This procedure is equivalent to Maximum Likelihood Estimation (MLE) if the responses are assumed to follow the asymmetric Laplace distribution (ALD), whose probability density function (p.d.f) is given by
\begin{eqnarray}
f_{i\boldsymbol{\beta}}(y) &=& \tau (1-\tau) \exp \left\{ -\rho_{\tau}(y-\mu_i^{\tau}) \right\},\ \mbox{ with } \mu^{\tau}_i ={\bf X}^{T}_i \boldsymbol{\beta} \mbox{ and } y\in(-\infty, \infty).\label{ALDpdf}
\end{eqnarray}
It is easy to check that $\mu^{\tau}_i$ is the $\tau^{th}$ quantile with respect to the p.d.f $f_{i\beta}(\cdot)$. 

The focus of this paper is a widely used Bayesian approach to the problem proposed by \cite{Yu_moyeed2001}, where the Bayesian posterior is obtained by assuming the likelihood  $f_{i \boldsymbol{\beta}}(\cdot)$ for $Y_i$, and  a prior $\Pi(\cdot)$ for $\boldsymbol{\beta}$. The approach is computationally attractive especially due to the location scale mixture normal representation of ALD (see \cite{Koz_Kob2011}), and is known to give posterior consistent estimates even if ALD is a misspecification (see \citealt{sriram.rvr.ghosh.2013}).  Therefore, the approach has been useful in many applications (e.g. \citealt{Yu_wage2005, Yue_rue2011, benoit2012, RahimYu2013, Waldmann2013}). 

The aim of this paper is to highlight and remedy an inferential limitation of the Bayesian quantile regression approach based on ALD, when the ALD model is possibly misspecified. From a classical Bayesian point of view, where the objective is to draw inference on  the``true" underlying quantile regression parameters ($\boldsymbol{\beta}_0$), it is desirable that the Bayesian inference coincides with the frequentist inference as the size of data increases.  For this, two ``frequentist" asymptotic properties need to hold, (i) Posterior consistency, i.e. the posterior asymptotically concentrates around the ``true" parameter value, and (ii) ``Coverage property", i.e.  the $100(1-\alpha)\%$ Bayesian credible sets asymptotically merge with the $100(1-\alpha)\%$ frequentist confidence sets around $\boldsymbol{\beta}^{M}_n$.  Even from a subjectivist Bayesian point of view,  which does not subscribe to the idea of a ``true parameter" value, \cite{diaconis_freedman} argue that violation of posterior consistency  is undesirable. Such violation would mean that two experts starting with different priors, can completely diverge on their opinions about the predictive distributions, even  as more data becomes available. Further, any small sample posterior inference based on a possibly misspecified likelihood is not straight forward to justify. In such cases, checking the large sample (asymptotic) properties is a way to avoid using an inappropriate likelihood.

The ALD is mainly used as a ``working likelihood" for Bayesian quantile inference. It is more often than not a misspecification of the true underlying likelihood. While posterior consistency still holds under suitable conditions (see \citealt{sriram.rvr.ghosh.2013}), we find that the ``coverage property" may not hold. An undesirable consequence would be that a narrow Bayesian credible interval could give a false sense of certainty about a parameter, which is an artifact of a misspecified likelihood rather than an actual gain from the Bayesian approach. This paper proposes a sandwich likelihood method to remedy this issue, while still leveraging the benefits of using ALD.

Section \ref{propmeth} describes the sandwich likelihood method. Supporting theoretical results and simulations are presented in Sections \ref{result} and \ref{simulation} respectively.
\section{The Sandwich Likelihood Method}
\label{propmeth}
Let $\tau\in(0,1)$ be fixed. Suppose $\{Y_i, \ i=1,2,\ldots,n\}$ are independent but non-identically distributed ($i.n.i.d$) with probability density function $Y_i\sim p_i(\cdot)$. Let the ``true" $\tau^{th}$ quantile of $Y_i$  be given by $Q_{\tau}({\bf X}_i)={\bf X}_i^{T}\boldsymbol{\beta}_0$, where ${\bf X}_i$ is a vector of $p-$dimensional non-random covariates.  We model the $\tau^{th}$ quantile as $Q_{\tau}({\bf X}_i)={\bf X}_i^{T}\boldsymbol{\beta}$ along with a proper prior $\Pi$ (with p.d.f $\pi$) for $\boldsymbol{\beta}$. We write the posterior distribution of $\boldsymbol{\beta}$ given the data $Y_1,Y_2, \ldots,Y_n$  as
 \begin{eqnarray}
 &&\Pi_{n}(\boldsymbol{\beta}\in B) = \frac{\int_{B}f_{\boldsymbol{\beta}}^{(n)}\pi(\boldsymbol{\beta})d\boldsymbol{\beta}}{\int_{\mathcal{R}^{d}}f_{\boldsymbol{\beta}}^{(n)}\pi(\boldsymbol{\beta})d\boldsymbol{\beta}}, \mbox{ where } f_{\boldsymbol{\beta}}^{(n)}= \prod_{i=1}^{n}f_{i\boldsymbol{\beta}}(Y_i). \label{posterior}
 \end{eqnarray}
The sandwich likelihood method is motivated from \cite{muller2013} and is based on two observations. 
\begin{enumerate} 
\item[] (a) As seen later in Theorem \ref{thmstep1}, the posterior distribution $\Pi_n(\cdot)$ of $\boldsymbol{\beta}$ is asymptotically equivalent to a normal distribution centred at $\boldsymbol{\beta}^{M}_n$ (as in equation(\ref{qrfreq}))  and covariance matrix $\frac{1}{n}V^{-1}$, where
\begin{eqnarray} 
V=\left(\lim_{n\rightarrow \ \infty} \frac{1}{n}\sum_{i=1}^{n}p_i({\bf X}^{T}_i \boldsymbol{\beta}_0){\bf X}_i {\bf X}^{T}_i \right). \label{varald}
\end{eqnarray}
\item[] (b)  It follows from \citealt{Koenker2005} (page 74) that $\boldsymbol{\beta}^{M}_n$ is asymptotically normal  with mean $\boldsymbol{\beta}_0$ and (``sandwich") covariance matrix $\frac{1}{n}\Sigma$, where 
\begin{eqnarray}
  \Sigma=\tau(1-\tau) V^{-1} S  V^{-1} \mbox{ with } \ S=\lim_{n\rightarrow \infty}S_n, \ S_n=\frac{1}{n}\sum_{i=1}^{n}{\bf X}_i{\bf X}^{T}_i \label{Smat}.
  \end{eqnarray}
\end{enumerate}
Since $V^{-1} \neq \Sigma$ in general,  (a) and (b) imply that  the credible intervals from $\Pi_n(\cdot)$ will not asymptotically match the normal frequentist confidence intervals, thus violating the ``coverage property" described in the introduction. 

Let $\boldsymbol{\tilde{\beta}}_n$ and  $\frac{1}{n}V^{-1}_n$  be the mean and covariance matrix for $\boldsymbol{\beta}$ under the posterior distribution $\Pi_n(\cdot)$. The ``Sandwich likelihood" method to remedy this issue can be described in two steps.
\begin{itemize}
\item[] {\bf Step 1.}(ALD step)  Using the posterior distribution  $\Pi_n(\cdot)$ purely based  on the ALD likelihood  $Y_{i}$ $ \sim f_{i\boldsymbol{\beta}}$ and prior $\Pi(\cdot)$, compute the  posterior mean $=\boldsymbol{\tilde{\beta}}_n$, and  covariance matrix $=\frac{1}{n}V_{n}^{-1}$.
\item[] {\bf Step 2.} (Sandwich Likelihood step) Define the ``Sandwich likelihood" as
 \begin{eqnarray}
&&g^{(n)}_{\boldsymbol{\beta}}=\frac{1}{ |\Sigma_n/n|^{\frac{1}{2}}}\cdot e^{-n( \boldsymbol{\tilde{\beta}}_n -\boldsymbol{\beta} )^{T} \Sigma_{n}^{-1}( \boldsymbol{\tilde{\beta}}_n-\boldsymbol{\beta} )} \label{sandwichL}\\
&&\mbox{ where } \Sigma_n = \tau(1-\tau)V_{n}^{-1} S_n V_{n}^{-1}, \label{Sigma}
\end{eqnarray}
~~~~~~~~~~~~~~and recompute a new posterior distribution for $\boldsymbol{\beta}$ as follows:
 \begin{eqnarray}
 &&\Pi^{prop}_{n}(\boldsymbol{\beta} \in B) = \frac{\int_{B}g^{(n)}_{\boldsymbol{\beta}} \pi(\boldsymbol{\beta})d\boldsymbol{\beta}}{\int_{\mathcal{R}^{d}}g^{(n)}_{\boldsymbol{\beta}}\pi(\boldsymbol{\beta})d\boldsymbol{\beta}}. \label{finalpost}
 \end{eqnarray}
\end{itemize}
Theorem \ref{thmstep2} in Section \ref{result} ensures that the credible sets based on the new posterior from Step 2  merge asymptotically with the frequentist confidence sets.  Here, we  make a few remarks.
\begin{remark}
Note that the proposed sandwich likelihood is based on the Bayesian posterior mean $\boldsymbol{\tilde{\beta}}_n$ from the ALD step (i.e. Step 1). We denote this proposed approach by ``SLBA". An alternative approach to sandwich likelihood in Step 2 is in the lines of \cite{muller2013}, got by using the classical quantile regression estimator $\boldsymbol{\beta}^{M}_n$ instead of $\boldsymbol{\tilde{\beta}}_n$ in equation (\ref{sandwichL}). We will denote this alternative by ``SLQR". While both these methods are asymptotically equivalent (see Lemma 2a), and work similarly for relatively flat priors or large sample sizes,  simulations in Section \ref{simulation} suggest that the proposed SLBA method may be more suited for small sample sizes with informative priors. \end{remark}
\begin{remark}
\label{R:computation} 
The Markov Chain Monte Carlo (MCMC) approaches for implementing Step 1 are now well known (e.g. see \citealt{Yue_rue2011}). Therefore, $\boldsymbol{\tilde{\beta}}_n$ and $\frac{1}{n}V_{n}^{-1}$ can be obtained as the mean and covariance matrix computed based on the MCMC simulations of $\boldsymbol{\beta}$ from Step 1. Consequently, $\Sigma_n$ (as in equation \ref{Sigma}) needed for Step 2 can be easily computed. 
\end{remark}
\begin{remark}
\label{R:postmeanvarcons}
Under the assumptions made in Section \ref{result}, Lemma \ref{centering2} shows that $\boldsymbol{\tilde{\beta}}_n$ and $V^{-1}_n$  are consistent for $\boldsymbol{\beta}_0$ and $V^{-1}$ respectively. Consequently, $\Sigma_n$ is consistent for $\Sigma$.  It is worth noting that estimation of the $\Sigma$ in the $i.n.i.d.$ case is in general a challenging problem (see section 3.4.2 of \citealt{Koenker2005}) since the true underlying densities $\{p_i(\cdot)\}_{i\geq 1}$ are not known. This method is a simple alternative.  
\end{remark}
 \section{Theoretical Results}
 \label{result}
 Recall that the true underlying p.d.f of $Y_i$ is $p_i$. By way of notation,  let $P^{(n)}$ denote the product probability $p_1\times p_2 \times\cdots p_n$ and  $P$ denote the infinite product probability $\prod_{i=1}^{\infty}p_i$. Let $P(\cdot)$ and $E\left[\cdot\right]$ denote the probability and expectation with respect to the true product probability. We define $Z_i:=Y_i-{\bf X}^{T}_i \boldsymbol{\beta}_0$ and note that $P(Z_i\leq 0|{\bf X}_i)=\tau$. 

Our Assumptions 1a, 2 to 4 are exactly those in  \cite{sriram.rvr.ghosh.2013} for posterior consistency. Assumption 1b  is additionally introduced to help ensure consistency of posterior mean and variance. 
\begin{itemize}
\item[{}] {\bf Assumption 1.}
\item[{}] (a) $\Pi$ is a proper prior with a bounded  and continuous p.d.f $\pi$, with $\pi(\boldsymbol{\beta}_0)>0$.
\item[] (b) $\int \|\boldsymbol{\beta}\|^2 \pi(\boldsymbol{\beta})d\boldsymbol{\beta}<\infty$.
\end{itemize}
 The second assumption requires that the covariates be bounded. 
 \begin{itemize}
\item[{}]{\bf Assumption 2.} $\exists \ M>0$ such that $\| {\bf X}_i\|\leq M, \ \forall \ i.$ 
\end{itemize}
The first part of the next assumption essentially says that the non-intercept covariates (after appropriate centering)  take values  in all quadrants of the Euclidean plane. In particular, this implies that they cannot be collinear. The second part of the assumption requires that the true underlying likelihood put positive mass around the true quantile, in particular ensuring that it is unique.
\begin{itemize}
\item[{}] {\bf Assumption 3.}
\item[{}] (a)  Let the first coordinate of ${\bf X}_i$ be identically 1 representing the intercept. After appropriate centering  of  the other co-ordinates,  $\exists \ \epsilon_0>0$ such that $\liminf_{n\rightarrow \infty}\frac{1}{n}\sum_{i=1}^{n}I_{{\bf X}_i \in D}>0$, $\forall$ $D\subset \mathcal{R}^{p}$ of the form $\{1\}\times U_2\times\cdots U_p$, where for some $j$, $U_j$ is either $=(\epsilon_0,\infty)$ or $(-\infty, -\epsilon_0)$ and for $k\neq j$, $U_k$ is either $(0,\infty)$ or $(-\infty, 0)$.
\item[{}] (b) For some $C>0$ and all sufficiently small $\Delta>0$, $P(0<Z_i<\Delta)>C\Delta$ and $P(-\Delta<Z_i<0)>C\Delta$ $\forall \ i$. 
\end{itemize}
Assumption 4 is a technical condition required for Strong Law of Large Numbers in the $i.n.i.d$ case.
\begin{itemize}
\item[{}] {\bf Assumption 4.}  $\limsup_{m\rightarrow \infty}\frac{1}{m}\sum_{i=1}^{m}E|Z_i|<\infty$ and  $\sum_{i=1}^{\infty}\frac{EZ_i^2}{i^{2}}<\infty$.
\end{itemize}
\begin{lemma}
\label{postcons}
Let Assumptions {\bf 1} to {\bf 4} hold, and $k\in\{0,1,2\}$. Then, for any sequence $M_n\rightarrow \infty$, 
\[E\left[\int_{\| \boldsymbol{\beta}-\boldsymbol{\beta}_0\|>\frac{M_n}{\sqrt{n}}}(\sqrt{n}\| \boldsymbol{\beta}-\boldsymbol{\beta}_0\|)^k d\Pi_n\right]\rightarrow 0.\]
\end{lemma}
For $k=0$, Lemma \ref{postcons} gives posterior consistency at $\sqrt{n}-$rate (the same conclusion as in Theorem 2b of \cite{sriram.rvr.ghosh.2013}). The results for $k=1$ and $k=2$ are useful in establishing the next lemma, which formalizes the fact that the classical estimator ($\boldsymbol{\beta}^{M}_n$) and the posterior mean from Step 1 (${\boldsymbol{\tilde{\beta}}}_n$) are asymptotically equivalent, and that the estimators ($V^{-1}_n$,$\Sigma_n$) are consistent for ($V^{-1}$,$\Sigma$).
\begin{lemma}
\label{centering2}
Under Assumption {\bf 1} to {\bf 4},
\begin{eqnarray*}
 (a) && \ \sqrt{n}({\boldsymbol{\beta}}^{M}_n- {\boldsymbol{\tilde{\beta}}}_n) \ \rightarrow \ 0 \mbox{ in probability } [P].\\
 (b) && \ V^{-1}_n \ \rightarrow \ V^{-1}, \mbox { and hence } \Sigma_n \rightarrow \Sigma\mbox{ in probability } [P].
\end{eqnarray*}
\end{lemma}
The asymptotic normality of the posterior is derived by applying the Bernstein-von-Mises theorem for misspecified models given by \cite{Kleijn_van2012}. For that, a key requirement apart from $\sqrt{n}-$ posterior consistency, is ``Local Asymptotic Normality"(LAN). For the ALD model on i.n.i.d data, the LAN property can be shown by making the following assumption on the boundedness and continuity of the true underlying densities $p_i$.
\begin{itemize}
\item[{}] {\bf Assumption 5.} 
\item[{}] For some $C, \eta >0$ and for all $\boldsymbol{\beta}$ in some small enough neighborhood of $\boldsymbol{\beta}_0$,  the p.d.fs $p_i$ satisfy:
\item[{}] (a) $  \{p_i({\bf X}^{T}_i\boldsymbol{\beta}_0 ), \ i\geq 1 \} \mbox{ are uniformly bounded away from } \infty.$
\item[{}] (b)$ \left\vert p_i({\bf X}^{T}_i\boldsymbol{\beta} )- p_i({\bf X}^{T}_i\boldsymbol{\beta}_0) \right\vert \ \leq \ C \left\vert{\bf X}^{T}_i\boldsymbol{\beta}-{\bf X}^{T}_i\boldsymbol{\beta}_0\right\vert^{\eta}  \ \forall \ i.$
\end{itemize}
\begin{lemma} [LAN property]
\label{lan}
Under Assumptions {\bf 2} and {\bf 5}, the LAN property holds, i.e., for any compact set $K\subset \mathcal{R}^{p}$,
\begin{eqnarray}
&&\sup_{\boldsymbol{\delta} \in K} \left\vert \log \frac{f^{(n)}_{\boldsymbol{\beta}_0+\frac{\boldsymbol{\delta}}{\sqrt{n}}}}{f^{(n)}_{\boldsymbol{\beta}_0}} - \boldsymbol{\delta}^T V \Delta_{n,\boldsymbol{\beta}_0} -\frac{1}{2}\boldsymbol{\delta}^{T}V\boldsymbol{\delta}\right\vert\rightarrow \ 0 \mbox{ in probability }\ [P].\\
&& \mbox{ where }, \Delta_{n,\boldsymbol{\beta}_0}= -V^{-1}\frac{1}{\sqrt{n}}\sum_{i=1}^{n}(\tau-I_{(Y_i\leq {\bf X}^{T}_i\boldsymbol{\beta}_0)}){\bf X}_i.
\end{eqnarray}
\end{lemma}
Proofs of Lemmas \ref{postcons}, \ref{centering2} and \ref{lan} are included in the Appendix and are essentially extensions of ideas from \cite{sriram.rvr.ghosh.2013}, \cite{Kleijn_van2012} and \cite{Koenker2005} respectively.   
The next lemma establishes the asymptotic connection between the posterior probability and the normal distribution. Let $\Phi(B, \boldsymbol{\mu}, \boldsymbol{\Sigma})$ denote the probability of a set $B$ under the multivariate normal distribution with mean $\boldsymbol{\mu}$ and covariance matrix $\boldsymbol{\Sigma}$.  Then, we have the following result.
\begin{lemma}
\label{asynormal}
Under Assumptions {\bf 1} to {\bf 5},
\begin{eqnarray}
(a) &&\sup_{B}\left\vert\Pi_n\left( \sqrt{n}(\boldsymbol{\beta}-\boldsymbol{\beta}_0)\in B \right)-\Phi\left(B, \Delta_{n, \boldsymbol{\beta}_0}, V^{-1} \right) \right\vert \ \rightarrow \ 0 \ \mbox{in probability } [P].\\
(b) && \sqrt{n}({\boldsymbol{\beta}}^{M}_n - \boldsymbol{\beta}_0)- \Delta_{n, \boldsymbol{\beta}_0} \ \rightarrow \ 0 \mbox{ in probability } [P].
\end{eqnarray}
\begin{proof}
Part(a) is an immediate consequence of Lemma \ref{postcons} (for $k=0$) and Lemma \ref{lan}, since these are precisely the conditions needed to apply Theorem 2.1 of \cite{Kleijn_van2012}. Part (b)  is  shown in \cite{Koenker2005} (see page 122, equation 4.4). 
\end{proof}
\end{lemma}
Our first theorem is immediate from Lemma \ref{asynormal} and formalizes Step 1.
\begin{theorem}
\label{thmstep1}
Let $\Pi_n $ be as in equations (\ref{posterior}). Then, under assumptions {\bf 1} to {\bf 5},
\begin{eqnarray*}
 && \sup_{B}\left\vert\Pi_n\left( \sqrt{n}(\boldsymbol{\beta}-\boldsymbol{\beta}_0)\in B \right)-\Phi\left(B, \sqrt{n}(\boldsymbol{\beta}^{M}_n - \boldsymbol{\beta}_0), V^{-1} \right) \right\vert \ \rightarrow \ 0 \ \mbox{in probability } [P].\\
\end{eqnarray*}
\end{theorem}
The next theorem formalizes Step 2 of the proposed Sandwich Likelihood approach.
\begin{theorem}
\label{thmstep2}
Let $\Pi_n $ and $\Pi^{prop}_n$ be as in equations (\ref{posterior}) and   (\ref{finalpost}). Then, under assumptions {\bf 1} to {\bf 5},
\begin{eqnarray*}
(a) && \sup_{B}\left\vert\Pi_n\left( \sqrt{n}(\boldsymbol{\beta}-\boldsymbol{\beta}_0)\in B \right)-\Phi\left(B, \sqrt{n}({\boldsymbol{\tilde{\beta}}}_n  - \boldsymbol{\beta}_0), V^{-1} \right) \right\vert \ \rightarrow \ 0 \ \mbox{in probability } [P].\\
(b) && \sup_{B}\left\vert\Pi^{prop}_n\left(\sqrt{n}(\boldsymbol{\beta}-\boldsymbol{\beta}_0)\in B\right)-\Phi\left(B,\sqrt{n}({\boldsymbol{\tilde{\beta}}}_n - \boldsymbol{\beta}_0), \Sigma\right) \right\vert \ \rightarrow \ 0 \ \mbox{in probability } [P].
\end{eqnarray*}
\begin{proof} Part (a) follows from Theorem \ref{thmstep1} and  Lemma \ref{centering2} (a).  To see part (b), first note that $\Sigma_n\rightarrow \Sigma$ in probability $[P]$ (by Lemma \ref{centering2}). Since $\sqrt{n}(\boldsymbol{\beta}^{M}_n - \boldsymbol{\beta}_0)$ is asymptotically normal (\cite{Koenker2005}), it is bounded in probability. Then, by Lemma \ref{centering2} (a), $\sqrt{n}(\boldsymbol{\tilde{\beta}}_n - \boldsymbol{\beta}_0)$ is also bounded in probability. Hence, $g^{(n)}_{\boldsymbol{\beta}}$ satisfies LAN property because
\begin{eqnarray*}
&&\sup_{\boldsymbol{\delta}\in\mathcal{K}}\left\vert\log\frac{g^{(n)}_{\boldsymbol{\beta}_0+\frac{\boldsymbol{\delta}}{\sqrt{n}}}}{g^{(n)}_{\boldsymbol{\beta}_0}}+ \boldsymbol{\delta}^{T}\Sigma^{-1}\sqrt{n}(\boldsymbol{\tilde{\beta}}_n - \boldsymbol{\beta}_0)- \frac{1}{2}\boldsymbol{\delta}^{T}\Sigma^{-1}\boldsymbol{\delta}\right\vert.\\
&&= \sup_{\boldsymbol{\delta}\in\mathcal{K}}\left\vert \boldsymbol{\delta}^{T}(\Sigma^{-1}-\Sigma_n^{-1})\sqrt{n}(\boldsymbol{\tilde{\beta}}_n - \boldsymbol{\beta}_0)- \frac{1}{2}\boldsymbol{\delta}^{T}(\Sigma^{-1}-\Sigma_n^{-1})\boldsymbol{\delta}\right\vert \rightarrow 0 \mbox{ in probability }[P].
\end{eqnarray*}
The result is immediate from Theorem 2.1  of \cite{Kleijn_van2012}, provided we show
\begin{eqnarray}
E\left[\Pi^{prop}_n\left(\sqrt{n}\|\boldsymbol{\beta}-\boldsymbol{\beta}_0\|>M_n \right)\right] \rightarrow \ 0 \mbox{ for any sequence } M_n \rightarrow \infty.\label{eq2show}
\end{eqnarray}
Let $B_n=\{\boldsymbol{\beta}: \ \sqrt{n}\|\boldsymbol{\beta}-\boldsymbol{\beta}_0\|>M_n\}$. Making a change of variable  $\sqrt{n}(\boldsymbol{\beta}-\boldsymbol{\tilde{\beta}}_n)=\boldsymbol{t}$, we can write
\begin{eqnarray*}
\Pi^{prop}_n (B_n)= \frac{ \int_{\mathcal{R}^d}\left(\frac{1}{\sqrt{2\pi \cdot det(\Sigma)}}\right)^d \cdot e^{-\frac{\boldsymbol{t}^{T}\Sigma^{-1}\boldsymbol{t}}{2}}\cdot I_{\left(\|\boldsymbol{\tilde{\beta}}_n+ \boldsymbol{t}/\sqrt{n}-\boldsymbol{\beta}_0 \|>M_n\right)}\cdot \pi(\boldsymbol{\tilde{\beta}}_n+\boldsymbol{t}/\sqrt{n}) d\boldsymbol{t}}{\int_{\mathcal{R}^d}\left(\frac{1}{\sqrt{2\pi \cdot det(\Sigma)}}\right)^d \cdot e^{-\frac{\boldsymbol{t}^{T}\Sigma^{-1}\boldsymbol{t}}{2}}\cdot \pi(\boldsymbol{\tilde{\beta}}_n+\boldsymbol{t}/\sqrt{n}) d\boldsymbol{t}}
\end{eqnarray*}
Since $\boldsymbol{\tilde{\beta}}_n\rightarrow \ \boldsymbol{\beta}_0 $ in probability $[P]$ ,  $M_n\rightarrow \ \infty$, and $\pi(\cdot)$ is bounded and continuous by Assumption 1, an application of Skorohod representation theorem for the sequence $\{\boldsymbol{\tilde{\beta}}_n\}_{n\geq 1}$ along with the dominated convergence theorem implies that the numerator of the above expression converges to zero and the denominator to  $\pi(\boldsymbol{\beta}_0)$. This in turn implies that $E\left[\Pi^{prop}_n\left(\sqrt{n}\|\boldsymbol{\beta}-\boldsymbol{\beta}_0\|>M_n \right)\right] \rightarrow \ 0$.
\end{proof}
\end{theorem}

\begin{remark}[On using ALD with a scale parameter]
In some applications,  one may carry out Bayesian quantile regresssion using an ALD that includes a scale parameter ($\sigma>0$), whose p.d.f is given by
\begin{eqnarray}
f_{i\boldsymbol{\beta}, \sigma}(y) &=& \frac{\tau (1-\tau)}{\sigma} \exp \left\{ -\frac{\rho_{\tau}(y-\mu_i^{\tau})}{\sigma} \right\},\ \mbox{ with } \mu^{\tau}_i ={\bf X}^{T}_i \boldsymbol{\beta} \mbox{ and } y\in(-\infty, \infty).\label{ALDpdfscale}
\end{eqnarray}
Here, either the scale is fixed (e.g. $\sigma=1$) or endowed with a prior. In the  latter case, the scale is essentially a nuisance parameter for Bayesian inference on $\boldsymbol{\beta}$. The proposed approach can be easily modified to incorporate the scale parameter.  Suppose the scale  is fixed at $\sigma=\sigma_0$. Then under Assumptions 1 to 5, it is easy to check that the conclusions of  Theorem \ref{thmstep2}(a)  will hold with $V^{-1}$ replaced by $\sigma_0 V^{-1}$. Accordingly, $\frac{1}{n}V_n^{-1}$ in Step 1 can be obtained as $\frac{1}{\sigma_0}$ times the estimated covariance matrix under $\Pi_n$. Step 2 would remain the same.

Suppose $\sigma$ is endowed with a prior on a compact interval $[\sigma_1, \sigma_2]$. Then, \cite{sriram.rvr.ghosh.2013} show under suitable conditions, that the posterior distribution  of $\sigma$ given $Y_1, Y_2, \ldots, Y_n$ would concentrate around a value $\sigma_0$ given by 
\begin{eqnarray}
&& \sigma_0= arg\max_{\sigma\in[\sigma_1, \sigma_2]} 
\log\left(\frac{\tau(1-\tau)}{\sigma}\right)  - \frac{C^{*}}{\sigma}, \label{sigma0} \\
\mbox{where, } && C^{*}=\lim_{m\rightarrow\infty}\frac{1}{m}\sum_{i=1}^{m}E(Z_i(\tau-I_{(Z_i\leq 0)})) \ \mbox{ and }  Z_i=Y_i- {\bf X}_i \boldsymbol{\beta}_0.\nonumber
\end{eqnarray}
In this case, it is reasonable to expect that under suitable conditions, conclusions of  Theorem \ref{thmstep2}(a)  will hold with $V^{-1}$ replaced by $\sigma_0 V^{-1}$, only now with $\sigma_0$ as in equation (\ref{sigma0}). It appears that a formal derivation of this result  requires the Bernstein-von-Mises theorem for misspecified models (as in \citealt{Kleijn_van2012}) to be developed in the presence of a nuisance parameter. Hence, a formal investigation is deferred to a future work. Here, one could first estimate $\sigma_0$ from Step 1 using the MCMC simulations  of the parameter $\sigma$ (e.g. posterior mean $\hat{\sigma}_0$), and then obtain $\frac{1}{n}V_n^{-1}$  as $\frac{1}{\hat{\sigma}_0}$ times the covariance matrix under $\Pi_n$. 
\end{remark}

\begin{table}[ht]
\caption{\label{tab1newsig} Comparison of methods when data size is large (N=2000). The numbers within parenthesis are coverage and interval length (COV in \%, LEN).}
\textsf{(a): N=2000, relatively flat prior on $\boldsymbol{\beta}$, Fixed ALD scale parameter at =1.
}
\centering
\resizebox{\textwidth}{!}{  
\begin{tabular}{|l|l|llll|llll|llll|}
  \hline
  & &  \multicolumn{3}{p{4cm}}{\centering $~~~~~~~~~~~~\alpha$}  & & \multicolumn{3}{p{4cm}}{\centering $~~~~~~~~~~~~\beta_1$}      & & \multicolumn{3}{p{4cm}}{\centering $~~~~~~~~~~~~\beta_2$} &  \\
  \hline
Model & $\tau$ & QR & ALD & SLQR & SLBA & QR & ALD & SLQR& SLBA & QR & ALD & SLQR & SLBA\\ 
     \hline
   \hline
1 & 0.25 & (94,0.43) & (98,0.55) & (95,0.42) & (96,0.42) & (94,0.13) & (98,0.17) & (94,0.13) & (95,0.13) & (94,0.26) & (98,0.34) & (93,0.26) & (92,0.26) \\ 
  2 & 0.25 & (92,0.18) & (100,0.36) & (92,0.18) & (94,0.18) & (94,0.06) & (100,0.11) & (94,0.06) & (95,0.06) & (94,0.11) & (100,0.22) & (94,0.11) & (96,0.11) \\ 
  3 & 0.25 & (96,2.54) & (72,1.31) & (96,2.52) & (95,2.52) & (98,0.92) & (64,0.44) & (94,0.92) & (94,0.92) & (96,2.17) & (68,0.96) & (96,2.15) & (96,2.15) \\ 
  4 & 0.25 & (95,2.76) & (72,1.36) & (95,2.72) & (95,2.72) & (95,1) & (68,0.46) & (96,0.99) & (97,0.99) & (94,2.4) & (57,1) & (92,2.33) & (92,2.33) \\ 
   \hline
   \hline
1 & 0.75 & (94,0.42) & (98,0.53) & (94,0.41) & (94,0.41) & (94,0.13) & (98,0.16) & (92,0.13) & (94,0.13) & (94,0.26) & (100,0.34) & (93,0.26) & (94,0.26) \\ 
  2 & 0.75 & (94,0.52) & (98,0.6) & (91,0.51) & (92,0.51) & (93,0.16) & (96,0.18) & (92,0.16) & (92,0.16) & (96,0.33) & (98,0.38) & (94,0.33) & (96,0.33) \\ 
  3 & 0.75 & (95,1.75) & (83,1.07) & (94,1.72) & (94,1.72) & (97,0.64) & (78,0.36) & (96,0.63) & (96,0.63) & (94,1.51) & (70,0.79) & (92,1.48) & (92,1.48) \\ 
  4 & 0.75 & (94,2.68) & (66,1.31) & (94,2.62) & (92,2.62) & (96,0.97) & (64,0.44) & (94,0.95) & (94,0.95) & (96,2.27) & (68,0.96) & (94,2.19) & (94,2.19) \\ 
   \hline
    \hline
\end{tabular}}
\textsf{(b): N=2000, relatively flat prior on $\boldsymbol{\beta}$, Inverse gamma prior on ALD scale parameter.
}
\centering
\resizebox{\textwidth}{!}{  
\begin{tabular}{|l|l|llll|llll|llll|}
  \hline
  & &  \multicolumn{3}{p{4cm}}{\centering $~~~~~~~~~~~~\alpha$}  & & \multicolumn{3}{p{4cm}}{\centering $~~~~~~~~~~~~\beta_1$}      & & \multicolumn{3}{p{4cm}}{\centering $~~~~~~~~~~~~\beta_2$} &  \\
  \hline
Model & $\tau$ & QR & ALD & SLQR & SLBA & QR & ALD & SLQR& SLBA & QR & ALD & SLQR & SLBA\\ 
     \hline
   \hline
1 & 0.25 & (94,0.41) & (86,0.3) & (95,0.41) & (94,0.41) & (96,0.13) & (86,0.09) & (96,0.12) & (95,0.12) & (98,0.26) & (83,0.19) & (96,0.26) & (96,0.26) \\ 
  2 & 0.25 & (96,0.18) & (96,0.16) & (96,0.18) & (97,0.18) & (97,0.05) & (96,0.05) & (96,0.05) & (97,0.05) & (91,0.11) & (90,0.1) & (91,0.11) & (92,0.11) \\ 
  3 & 0.25 & (96,2.44) & (96,2.17) & (96,2.4) & (98,2.4) & (96,0.88) & (92,0.72) & (96,0.87) & (97,0.87) & (93,2.11) & (88,1.62) & (92,2.08) & (94,2.08) \\ 
  4 & 0.25 & (94,2.71) & (86,2.11) & (91,2.63) & (94,2.63) & (94,0.98) & (86,0.7) & (92,0.96) & (92,0.96) & (95,2.36) & (80,1.59) & (92,2.32) & (92,2.32) \\ 
   \hline
   \hline
   1 & 0.75 & (95,0.42) & (84,0.3) & (93,0.41) & (94,0.41) & (96,0.13) & (82,0.09) & (94,0.13) & (94,0.13) & (95,0.27) & (84,0.19) & (96,0.27) & (96,0.27) \\ 
  2 & 0.75 & (96,0.53) & (80,0.35) & (94,0.52) & (94,0.52) & (96,0.16) & (80,0.11) & (94,0.16) & (93,0.16) & (95,0.34) & (80,0.22) & (92,0.33) & (94,0.33) \\ 
 3 & 0.75 & (92,1.68) & (80,1.25) & (90,1.65) & (90,1.65) & (89,0.62) & (76,0.42) & (89,0.61) & (88,0.61) & (94,1.54) & (78,0.97) & (93,1.5) & (94,1.5) \\ 
  4 & 0.75 & (94,2.63) & (91,2.07) & (94,2.58) & (94,2.58) & (96,0.98) & (84,0.7) & (94,0.96) & (94,0.96) & (92,2.36) & (80,1.59) & (90,2.33) & (90,2.33) \\ 
  \hline
  \hline
\end{tabular}}
\textsf{(c): N=2000, Informative prior on $\boldsymbol{\beta}$, Fixed ALD scale parameter at =1.
}
\centering
\resizebox{\textwidth}{!}{  
\begin{tabular}{|l|l|llll|llll|llll|}
  \hline
  & &  \multicolumn{3}{p{4cm}}{\centering $~~~~~~~~~~~~\alpha$}  & & \multicolumn{3}{p{4cm}}{\centering $~~~~~~~~~~~~\beta_1$}      & & \multicolumn{3}{p{4cm}}{\centering $~~~~~~~~~~~~\beta_2$} &  \\
  \hline
Model & $\tau$ & QR & ALD & SLQR & SLBA & QR & ALD & SLQR& SLBA & QR & ALD & SLQR & SLBA\\ 
     \hline
   \hline
1 & 0.25 & (96,0.42) & (100,0.53) & (98,0.41) & (96,0.41) & (96,0.13) & (100,0.16) & (96,0.13) & (96,0.13) & (96,0.26) & (100,0.34) & (95,0.26) & (96,0.26) \\ 
  2 & 0.25 & (94,0.18) & (100,0.35) & (95,0.17) & (98,0.17) & (96,0.06) & (100,0.11) & (96,0.05) & (96,0.05) & (95,0.11) & (100,0.22) & (96,0.11) & (95,0.11) \\ 
  3 & 0.25 & (98,2.43) & (78,1.19) & (98,1.82) & (99,1.82) & (96,0.88) & (69,0.4) & (96,0.69) & (97,0.69) & (96,2.12) & (68,0.91) & (96,1.72) & (96,1.72) \\ 
  4 & 0.25 & (96,2.68) & (76,1.24) & (96,1.94) & (98,1.94) & (97,0.98) & (72,0.42) & (95,0.74) & (96,0.74) & (94,2.36) & (68,0.95) & (92,1.82) & (96,1.82) \\ 
   \hline
   \hline
1 & 0.75 & (94,0.42) & (98,0.53) & (93,0.4) & (95,0.4) & (94,0.13) & (98,0.16) & (92,0.12) & (94,0.12) & (94,0.26) & (100,0.34) & (93,0.26) & (93,0.26) \\ 
  2 & 0.75 & (94,0.52) & (98,0.59) & (90,0.49) & (92,0.49) & (93,0.16) & (96,0.18) & (91,0.15) & (94,0.15) & (96,0.33) & (98,0.38) & (95,0.33) & (97,0.33) \\ 
  3 & 0.75 & (95,1.75) & (84,1.02) & (94,1.45) & (95,1.45) & (97,0.64) & (78,0.34) & (96,0.54) & (96,0.54) & (94,1.51) & (72,0.78) & (94,1.33) & (94,1.33) \\ 
  4 & 0.75 & (94,2.68) & (70,1.24) & (94,1.93) & (96,1.93) & (96,0.97) & (66,0.41) & (93,0.74) & (95,0.74) & (96,2.27) & (68,0.93) & (96,1.78) & (96,1.78) \\ 
   \hline
    \hline
\end{tabular}}

\textsf{(d): N=2000, Informative prior on $\boldsymbol{\beta}$, Inverse gamma prior on ALD scale parameter.
}
\centering
\resizebox{\textwidth}{!}{  
\begin{tabular}{|l|l|llll|llll|llll|}
  \hline
  & &  \multicolumn{3}{p{4cm}}{\centering $~~~~~~~~~~~~\alpha$}  & & \multicolumn{3}{p{4cm}}{\centering $~~~~~~~~~~~~\beta_1$}      & & \multicolumn{3}{p{4cm}}{\centering $~~~~~~~~~~~~\beta_2$} &  \\
  \hline
Model & $\tau$ & QR & ALD & SLQR & SLBA & QR & ALD & SLQR& SLBA & QR & ALD & SLQR & SLBA\\ 
     \hline
   \hline
   1 & 0.25 & (96,0.42) & (86,0.29) & (95,0.4) & (96,0.4) & (96,0.13) & (88,0.09) & (94,0.12) & (96,0.12) & (94,0.27) & (86,0.19) & (94,0.26) & (94,0.26) \\ 
  2 & 0.25 & (96,0.17) & (95,0.16) & (96,0.17) & (94,0.17) & (94,0.05) & (92,0.05) & (94,0.05) & (93,0.05) & (94,0.12) & (94,0.1) & (94,0.11) & (94,0.11) \\ 
  3 & 0.25 & (94,2.39) & (96,1.85) & (88,1.59) & (97,1.59) & (95,0.88) & (94,0.63) & (90,0.62) & (96,0.62) & (96,2.2) & (94,1.53) & (94,1.64) & (98,1.64) \\ 
  4 & 0.25 & (94,2.71) & (91,1.84) & (88,1.76) & (98,1.76) & (92,0.98) & (86,0.62) & (90,0.68) & (96,0.68) & (95,2.44) & (85,1.51) & (93,1.81) & (96,1.81) \\ 
   \hline
   \hline
1 & 0.75 & (95,0.42) & (88,0.3) & (94,0.41) & (94,0.41) & (96,0.13) & (87,0.09) & (96,0.13) & (96,0.13) & (92,0.27) & (86,0.19) & (92,0.26) & (92,0.26) \\ 
  2 & 0.75 & (96,0.54) & (82,0.36) & (96,0.52) & (96,0.52) & (96,0.16) & (84,0.11) & (97,0.16) & (97,0.16) & (96,0.34) & (84,0.22) & (96,0.34) & (96,0.34) \\ 
  3 & 0.75 & (98,1.73) & (91,1.22) & (94,1.4) & (98,1.4) & (96,0.63) & (84,0.41) & (92,0.52) & (96,0.52) & (94,1.56) & (78,0.95) & (90,1.33) & (92,1.33) \\ 
  4 & 0.75 & (94,2.73) & (94,1.86) & (89,1.75) & (96,1.75) & (94,0.99) & (88,0.63) & (89,0.69) & (98,0.69) & (94,2.38) & (84,1.47) & (94,1.74) & (98,1.74) \\ 
   \hline
    \hline
\end{tabular}}
\end{table}

\section{Simulation Study}
\label{simulation}
Here, we study the performance of the proposed sandwich likelihood method (SLBA) by simulating data from different ``true" underlying models. We work with two covariates $X_1$ and $X_2$, simulated from $N(3,1)$ (truncated between 1 and 1000) and $Bernoulli(0.3)$ respectively. For each simulated model, we ensure that the true $\tau^{th}$-quantile is given by $q_{\tau}({\bf X})=(1+2X_1+3X_2)$. The simulated models are described below. The first two models are a location shifted normal and a location shifted gamma respectively, the third is a  scaled gamma, and the fourth model a location shifted and scaled normal. 
\begin{itemize}
\item[{}] {\bf Model 1.} $Y =q_{\tau}({\bf X}) + \epsilon$, where $\epsilon=Z-\rho_{\tau}$, $Z \sim N(0,1)$ and $\rho_{\tau}=\tau^{th}$ quantile of $N(0,1)$.
\item[{}] {\bf Model 2.} $ Y=1+2X_{1}+3 X_{2}-\rho_{\tau}+e$, where  $e\sim Gamma(shape=1,scale=1), \ \rho_{\tau}$ is the $\tau^{th}$ quantile of $Gamma(1,1)$
\item[{}] {\bf Model 3.} $Y \sim Gamma\left(shape=2, scale=\frac{\rho_{\tau}}{q_{\tau}({\bf X})}\right)$, where  $\rho_{\tau}=\tau^{th}$ quantile of $Gamma(2,1)$.
\item[{}]{\bf Model 4.}  $ Y\sim N(1+2 X_{1}+3 X_{2}-\rho_{\tau}|1+2 X_{1}+3 X_{2}|,|1+2 X_{1}+3 X_{2}|^{2})$, where $\rho_{\tau}$ is the $\tau^{th}$ quantile of $N(0,1)$.
 \end{itemize} 
The specified  model for the $\tau^{th}$ quantile is  $Q_{\tau}({\bf X})=\alpha+\beta_1 X_1 + \beta_2 X_2$. 
We present results for different quantiles ($\tau \in \{ 0.25, .75\}$), different sample sizes ($N \in \{50, 2000\}$) and analyze with respect to a relatively flat prior (i.e. a product of three N(0,100) distributions), as well as an informative prior for the parameters $(\alpha, \beta_1, \beta_2)$ (i.e., a product of N(.9,1), N(2.1,1) and N(2.9,1)). In addition, we present two scenarios for the scale parameter of ALD, (i)$\sigma=1$  and (ii) $\sigma\sim$ Gamma prior with mean=100 and variance=1000. The conclusions reached were similar for other scenarios involving values of $\tau\in\{0.05,.5,.95\}$, $\sigma\in\{1/2, 2\}$, and $N\in\{100,500\}$. However, these scenarios are not shown here in the interest of conciseness.

Recall that the proposed sandwich likelihood (denoted SLBA) is based on the posterior mean $\boldsymbol{\tilde{\beta}}_n$ from Step 1. We compare this with the alternative method of using the classical  estimator $\boldsymbol{\beta}^M_n$ (denoted by SLQR). We also compare with the frequentist quantile regression (denoted QR) and the Bayesian approach based purely on ALD, which is same as Step 1 (denoted ALD). For the QR method, we use the bootstrap method for computing confidence intervals and for the other methods we compute the posterior Bayesian credible intervals using 1000 MCMC simulations after a burn-in of 2000 simulations. The methods are compared with respect to the coverage property of the 95\% confidence/credible interval (denoted by COV),  and the length of the confidence interval (denoted by LEN).  The coverage (COV) is computed by repeating the above simulation 200 times and then calculating the percentage of times the 95\% confidence/credible intervals contain the true value. Similarly, the length (LEN) is computed as the average interval length across the 200 repetitions.

Table \ref{tab1newsig} compares the methods across different scenarios when the sample size is large ($N=2000$), and Table \ref{tab1newsign50} does the comparison when sample size is small ($N=50$). The sub-tables (a) and (b) are for a relatively flat prior on $\boldsymbol{\beta}$, and sub-tables (c) and (d) are for an informative prior. Further, sub-table (a) and (c) fix the ALD scale parameter at 1, whereas sub-tables (b) and (d) carry out the analysis by considering a gamma prior on the ALD scale parameter. 

We can make the following observations from Table \ref{tab1newsig}. For large $N$(=2000), it is desirable that Bayesian and the classical inferences merge asymptotically.  Here, as expected from our results, the coverages and length of intervals from both the sandwich likelihood methods viz., the proposed SLBA and the alternative SLQR,  as well as the classical QR methods are close to each other. SLBA is closer to SLQR for relatively flat priors, and performs slightly better with an informative prior. However, compared to these methods, the coverages from the ALD method in Tables 1(a) and (c) (i.e. scale=1) are not as close to 95\%. For the simpler models 1 and 2 with $i.i.d.$ errors, coverages under ALD are consistently but slightly higher than 95\%. For more complex models 3 and 4 with $i.n.i.d$ errors, coverages are way below 95\%. So, for more complex models, the inadequacy of coverage under ALD is more pronounced.  Such an issue with ALD can be partly addressed by allowing some flexibility in the ALD scale parameter.  This is checked in Tables 1(b) and 1(d), where a prior is assumed on $\sigma$ and the coverages from the ALD method are seen to improve for models 3 and 4. Since ALD is still a misspecificaton, the issue does not fully go away ( e.g. as seen in the coverages for $\beta_1$ in Table 1(d) for $\tau=.75$).  Finally, since the sample size is large, the results are similar for a relatively flat prior  and for an informative prior. In summary, the simulation results in Table \ref{tab1newsig} are supportive of the asymptotic results in Section \ref{result}. They also highlight the fact that using Bayesian inference based purely on  ALD  could be more misleading especially when the true data generating likelihood is complex.

\begin{table}[ht]
\caption{\label{tab1newsign50} Comparison of methods when data size is small (N=50). The numbers within parenthesis are coverage and interval length (COV in \%, LEN).}
\textsf{(a): N=50, relatively flat prior on $\boldsymbol{\beta}$, Fixed ALD scale parameter at =1.
}
\centering
\resizebox{\textwidth}{!}{  
\begin{tabular}{|l|l|llll|llll|llll|}
  \hline
  & &  \multicolumn{3}{p{4cm}}{\centering $~~~~~~~~~~~~\alpha$}  & & \multicolumn{3}{p{4cm}}{\centering $~~~~~~~~~~~~\beta_1$}      & & \multicolumn{3}{p{4cm}}{\centering $~~~~~~~~~~~~\beta_2$} &  \\
  \hline
Model & $\tau$ & QR & ALD & SLQR & SLBA & QR & ALD & SLQR& SLBA & QR & ALD & SLQR & SLBA\\ 
     \hline
   \hline
1 & 0.25 & (92,2.78) & (100,3.53) & (95,2.91) & (98,2.91) & (94,0.88) & (100,1.12) & (96,0.93) & (97,0.93) & (94,1.92) & (100,2.43) & (96,2.06) & (98,2.06) \\ 
  2 & 0.25 & (97,1.26) & (100,2.72) & (100,1.72) & (100,1.72) & (98,0.4) & (100,0.87) & (100,0.56) & (100,0.56) & (96,0.88) & (100,1.88) & (100,1.23) & (100,1.23) \\ 
  3 & 0.25 & (96,16.5) & (75,7.37) & (94,12.59) & (95,12.59) & (97,6.11) & (71,2.55) & (92,4.81) & (94,4.81) & (96,14.66) & (67,5.92) & (90,11.98) & (92,11.98) \\ 
  4 & 0.25 & (96,18.05) & (78,7.7) & (92,13.64) & (94,13.64) & (96,6.72) & (68,2.68) & (92,5.29) & (92,5.29) & (96,15.78) & (66,6.23) & (91,13.09) & (93,13.09) \\ 
   \hline
   \hline
   1 & 0.75 & (94,2.67) & (100,3.32) & (95,2.73) & (96,2.73) & (96,0.78) & (100,0.98) & (96,0.82) & (98,0.82) & (94,1.68) & (100,2.11) & (96,1.74) & (99,1.74) \\ 
  2 & 0.75 & (98,3.28) & (100,3.55) & (97,3.17) & (99,3.17) & (97,0.94) & (100,1.03) & (98,0.92) & (99,0.92) & (95,2.19) & (99,2.34) & (96,2.17) & (98,2.17) \\ 
  3 & 0.75 & (96,12.84) & (82,6.57) & (94,11.05) & (95,11.05) & (94,4.43) & (74,2.1) & (91,3.88) & (92,3.88) & (98,9.17) & (72,4.56) & (96,8.24) & (95,8.24) \\ 
  4 & 0.75 & (98,20.55) & (74,8.13) & (95,15.6) & (96,15.6) & (96,6.97) & (67,2.59) & (90,5.44) & (91,5.44) & (98,14.16) & (64,5.55) & (92,11.71) & (95,11.71) \\ 
   \hline
    \hline
\end{tabular}}
\textsf{(b): N=50, relatively flat prior on $\boldsymbol{\beta}$, Inverse gamma prior on ALD scale parameter.
}
\centering
\resizebox{\textwidth}{!}{  
\begin{tabular}{|l|l|llll|llll|llll|}
  \hline
  & &  \multicolumn{3}{p{4cm}}{\centering $~~~~~~~~~~~~\alpha$}  & & \multicolumn{3}{p{4cm}}{\centering $~~~~~~~~~~~~\beta_1$}      & & \multicolumn{3}{p{4cm}}{\centering $~~~~~~~~~~~~\beta_2$} &  \\
  \hline
Model & $\tau$ & QR & ALD & SLQR & SLBA & QR & ALD & SLQR& SLBA & QR & ALD & SLQR & SLBA\\ 
  1 & 0.25 & (97,3.37) & (80,2.01) & (93,3.25) & (92,3.25) & (96,1.05) & (80,0.62) & (92,1) & (92,1) & (97,1.82) & (80,1.09) & (95,1.72) & (96,1.72) \\ 
  2 & 0.25 & (98,1.49) & (94,1.12) & (96,1.44) & (98,1.44) & (99,0.47) & (96,0.35) & (97,0.45) & (97,0.45) & (96,0.84) & (92,0.61) & (93,0.77) & (97,0.77) \\ 
  3 & 0.25 & (98,21.96) & (94,14.06) & (93,15.35) & (97,15.35) & (98,7.43) & (94,4.6) & (93,5.34) & (97,5.34) & (96,14.71) & (86,8.98) & (92,11.86) & (96,11.86) \\ 
  4 & 0.25 & (98,22.92) & (92,13.3) & (96,16.2) & (98,16.2) & (98,7.74) & (86,4.35) & (94,5.67) & (96,5.67) & (94,15.44) & (86,8.7) & (93,12.94) & (94,12.94) \\    \hline
   \hline
   1 & 0.75 & (94,3.01) & (84,1.82) & (91,2.9) & (93,2.9) & (94,0.94) & (83,0.56) & (90,0.92) & (93,0.92) & (96,2.16) & (90,1.28) & (96,2.08) & (96,2.08) \\ 
  2 & 0.75 & (94,3.63) & (80,2.05) & (89,3.5) & (92,3.5) & (93,1.11) & (82,0.63) & (91,1.08) & (92,1.08) & (94,2.64) & (82,1.42) & (90,2.44) & (93,2.44) \\ 
  3 & 0.75 & (93,12.81) & (86,7.57) & (92,10.97) & (94,10.97) & (94,4.67) & (82,2.55) & (92,4.06) & (94,4.06) & (93,13.49) & (73,6.38) & (90,10.93) & (92,10.93) \\ 
  4 & 0.75 & (96,19.67) & (88,12.13) & (91,15.08) & (96,15.08) & (92,7.01) & (80,4.04) & (90,5.47) & (93,5.47) & (94,20.65) & (80,10.44) & (91,15.54) & (92,15.54) \\ 
   \hline
   \hline

  \hline
  \hline
\end{tabular}}
\textsf{(c): N=50, Informative prior on $\boldsymbol{\beta}$, Fixed ALD scale parameter at =1.
}
\centering
\resizebox{\textwidth}{!}{  
\begin{tabular}{|l|l|llll|llll|llll|}
  \hline
  & &  \multicolumn{3}{p{4cm}}{\centering $~~~~~~~~~~~~\alpha$}  & & \multicolumn{3}{p{4cm}}{\centering $~~~~~~~~~~~~\beta_1$}      & & \multicolumn{3}{p{4cm}}{\centering $~~~~~~~~~~~~\beta_2$} &  \\
  \hline
Model & $\tau$ & QR & ALD & SLQR & SLBA & QR & ALD & SLQR& SLBA & QR & ALD & SLQR & SLBA\\ 
     \hline
   \hline
1 & 0.25 & (96,2.69) & (100,2.46) & (84,1.38) & (98,1.38) & (96,0.79) & (100,0.76) & (86,0.48) & (98,0.48) & (96,1.7) & (100,1.79) & (94,1.19) & (98,1.19) \\ 
  2 & 0.25 & (96,1.18) & (100,2.04) & (98,0.99) & (100,0.99) & (96,0.35) & (100,0.63) & (98,0.33) & (100,0.33) & (95,0.75) & (100,1.4) & (98,0.74) & (99,0.74) \\ 
  3 & 0.25 & (98,18.89) & (98,3.35) & (36,2.28) & (100,2.28) & (96,6.39) & (79,1.31) & (74,1.61) & (92,1.61) & (96,12.68) & (87,3.09) & (66,2.74) & (98,2.74) \\ 
  4 & 0.25 & (96,19.72) & (98,3.37) & (30,2.3) & (99,2.3) & (92,6.71) & (78,1.33) & (72,1.69) & (91,1.69) & (96,13.81) & (86,3.14) & (64,2.79) & (98,2.79) \\ 
   \hline
   \hline
1 & 0.75 & (94,2.67) & (100,2.45) & (78,1.38) & (98,1.38) & (96,0.78) & (100,0.76) & (84,0.48) & (98,0.48) & (94,1.68) & (100,1.8) & (90,1.2) & (98,1.2) \\ 
  2 & 0.75 & (98,3.28) & (100,2.55) & (82,1.48) & (100,1.48) & (97,0.94) & (100,0.79) & (86,0.52) & (100,0.52) & (95,2.19) & (100,1.93) & (91,1.37) & (98,1.37) \\ 
  3 & 0.75 & (96,12.84) & (96,3.23) & (44,2.17) & (97,2.17) & (94,4.43) & (88,1.23) & (76,1.41) & (92,1.41) & (98,9.17) & (90,2.92) & (72,2.57) & (98,2.57) \\ 
  4 & 0.75 & (98,20.55) & (98,3.4) & (34,2.32) & (98,2.32) & (96,6.97) & (77,1.34) & (72,1.69) & (88,1.69) & (98,14.16) & (88,3.15) & (65,2.79) & (100,2.79) \\ 
   \hline
    \hline
\end{tabular}}

\textsf{(d): N=50, Informative prior on $\boldsymbol{\beta}$, Inverse gamma prior on ALD scale parameter.
}
\centering
\resizebox{\textwidth}{!}{  
\begin{tabular}{|l|l|llll|llll|llll|}
  \hline
  & &  \multicolumn{3}{p{4cm}}{\centering $~~~~~~~~~~~~\alpha$}  & & \multicolumn{3}{p{4cm}}{\centering $~~~~~~~~~~~~\beta_1$}      & & \multicolumn{3}{p{4cm}}{\centering $~~~~~~~~~~~~\beta_2$} &  \\
  \hline
Model & $\tau$ & QR & ALD & SLQR & SLBA & QR & ALD & SLQR& SLBA & QR & ALD & SLQR & SLBA\\ 
     \hline
   \hline
   1 & 0.25 & (94,2.68) & (86,1.48) & (92,1.82) & (95,1.82) & (95,0.78) & (84,0.43) & (92,0.56) & (94,0.56) & (96,1.71) & (87,0.98) & (94,1.36) & (95,1.36) \\ 
  2 & 0.25 & (98,1.15) & (94,0.85) & (95,0.97) & (98,0.97) & (98,0.34) & (94,0.25) & (96,0.29) & (96,0.29) & (96,0.76) & (93,0.55) & (92,0.66) & (96,0.66) \\ 
  3 & 0.25 & (98,19.18) & (100,3.6) & (12,1.27) & (98,1.27) & (95,6.5) & (98,1.67) & (52,1.54) & (94,1.54) & (98,13.5) & (100,3.57) & (28,1.87) & (98,1.87) \\ 
  4 & 0.25 & (98,19.58) & (100,3.57) & (17,1.44) & (100,1.44) & (96,6.7) & (94,1.63) & (54,1.62) & (94,1.62) & (96,14.52) & (100,3.52) & (34,2.05) & (98,2.05) \\ 
   \hline
   \hline
   1 & 0.75 & (92,2.7) & (81,1.5) & (92,1.84) & (92,1.84) & (94,0.82) & (82,0.46) & (90,0.59) & (96,0.59) & (96,1.77) & (82,1.03) & (94,1.41) & (96,1.41) \\ 
  2 & 0.75 & (94,3.34) & (85,1.67) & (93,2.02) & (96,2.02) & (95,1.02) & (86,0.52) & (94,0.68) & (96,0.68) & (96,2.31) & (84,1.2) & (95,1.68) & (96,1.68) \\ 
  3 & 0.75 & (96,11.39) & (100,3.27) & (38,1.94) & (100,1.94) & (94,4) & (89,1.27) & (66,1.31) & (92,1.31) & (94,10.29) & (96,3.15) & (68,2.46) & (100,2.46) \\ 
  4 & 0.75 & (96,18.73) & (100,3.56) & (16,1.42) & (98,1.42) & (95,6.66) & (90,1.62) & (51,1.61) & (92,1.61) & (98,15.68) & (100,3.58) & (32,2.01) & (99,2.01) \\ 
   \hline
    \hline
\end{tabular}}
\end{table}
To further help demonstrate the usefulness of SLBA, Table \ref{tab1newsign50} shows the simulation results for a small sample size $N=50$. When the prior is relatively flat, as in sub-tables (a) and (b), the observations made in the previous paragraph still more or less hold.  However, when we use an informative prior, as in sub tables (c) and (d), differences start to show. Unlike sub tables (a) and  (b), where the lengths of credible intervals for the QR and SLBA methods are similar, the SLBA method has much smaller intervals than QR in sub-tables (c) and (d) , while still retaining coverages around 95\%. This should not be surprising since the QR method is a classical approach and does not utilize the prior information, whereas the SLBA method utilizes the informative prior. In contrast, the SLQR method is seen to perform poorly. This is because SLQR leads to intervals of similar length as SLBA, but is centered at the classical estimator ($\boldsymbol{\beta}^M_n$). Since $\boldsymbol{\beta}^M_n$ does not use the informative prior, this centering can be inaccurate. The ALD method performs better with an informative prior, but SLBA continues  to have better coverage.

In summary,  the proposed SLBA method works better than the SLQR, ALD and QR methods across all scenarios. Further, while SLBA and SLQR are asymptotically equivalent and perform similarly in the case of large sample sizes or relatively flat priors,  the SLBA method is seen to consistently work better and more pronouncedly so for small sample sizes and informative priors.
 \FloatBarrier

\section*{Appendix}

\begin{proof}[{\bf \underline{Proof of Lemma 1}}]
The proof is by extending ideas from \cite{sriram.rvr.ghosh.2013}. A sketch is provided here. Let $\Delta_n=\frac{M_n}{\sqrt{n}}$ and $k\in\{0,1,2\}$. Following arguments leading up to Lemma 5 of their paper, a compact set $G\subseteq \{\|\boldsymbol{\beta}-\boldsymbol{\beta}_0\|\leq M_0\}$ exists such that for some $u>0$ and sufficiently large $n$,
\begin{equation}
\label{argongcomp}
\int_{G^c}(\sqrt{n}\|\boldsymbol{\beta}-\boldsymbol{\beta}_0 \|)^k \frac{f_{\boldsymbol{\beta}}^{(n)}}{f_{\boldsymbol{\beta_0}}^{(n)}}d\Pi(\boldsymbol{\beta})< n^{\frac{k}{2}}e^{-nu}\int_{G^c}\|\boldsymbol{\beta}-\boldsymbol{\beta}_0 \|^k \pi(\boldsymbol{\beta})d\boldsymbol{\beta}.
\end{equation}
By Assumption 1b, the integral on the right hand side is finite, and hence goes to zero as $n\rightarrow \infty$. Along with the fact that for any $\epsilon>0$, $e^{n\epsilon}\int\frac{f_{\boldsymbol{\beta}}^{(n)}}{f_{\boldsymbol{\beta_0}}^{(n)}}d\Pi(\boldsymbol{\beta})\rightarrow \ \infty \ a.s. [P]$, it will then follow that 
\begin{equation}
\label{ExpGcomp}
E\left[\int_{G^c }(\sqrt{n}\|\boldsymbol{\beta}-\boldsymbol{\beta}_0 \|)^k d\Pi_n(\boldsymbol{\beta})\right] \rightarrow \ 0.
\end{equation}

Now, for any $\epsilon_1>0$ (to be chosen later), using arguments leading up to the proof of Theorem 1 of \cite{sriram.rvr.ghosh.2013}, there exists $0<d<1$ and some positive constants $C_1', C'_2$ such that
\begin{eqnarray*}
&& E\left[ \int_{G\cap\{\|\boldsymbol{\beta}-\boldsymbol{\beta}_0 \|>\epsilon_1\}}(\sqrt{n}\|\boldsymbol{\beta}-\boldsymbol{\beta}_0 \|)^k d\Pi_{n}(\boldsymbol{\beta})\right] \leq n^{\frac{k}{2}}M^{k}_0 E\left[ \Pi_n\left(G\cap\{\|\boldsymbol{\beta}-\boldsymbol{\beta}_0 \|>\epsilon_1\} \right)\right]\nonumber\\
&& \leq n^{\frac{k}{2}}M^{k}_0 E\left[ \left(\Pi_n\left(G\cap\{\|\boldsymbol{\beta}-\boldsymbol{\beta}_0 \|>\epsilon_1\} \right)\right)^d\right]\leq C' n^{\frac{k}{2}}M^{k}_0 e^{-n\epsilon^2_1C'_2}
\end{eqnarray*}
Since the right hand side of the above inequality converges to zero as $n\rightarrow  \ \infty$, we get
\begin{equation}
\label{ExpGeps}
E\left[ \int_{G\cap\{\|\boldsymbol{\beta}-\boldsymbol{\beta}_0 \|>\epsilon_1\}}(\sqrt{n}\|\boldsymbol{\beta}-\boldsymbol{\beta}_0 \|)^k d\Pi_{n}(\boldsymbol{\beta})\right] \rightarrow \ 0.
\end{equation}
By Assumption 2 and Lemma 1 of \cite{sriram.rvr.ghosh.2013}, there exists a constant $C_3>0$ such that 
\[\forall \ \boldsymbol{\beta}: \|\boldsymbol{\beta}-\boldsymbol{\beta}_0 \|<\epsilon_1:= \frac{1}{4C_3}, \mbox{ we have }\  \left\vert\log\frac{f_{\boldsymbol{\beta}}}{f_{\boldsymbol{\beta_0}}}\right\vert\leq C_3 \|\boldsymbol{\beta}-\boldsymbol{\beta}_0 \|< \frac{1}{2} .\]
Since for $t<1/2$, $e^{t}-1<\frac{t}{1-t}<2t$, and using the fact that $E\left[\log\frac{f_{\boldsymbol{\beta}_0}}{f_{\boldsymbol{\beta}}}\right]>0$, we have
\[\forall \ \boldsymbol{\beta}\in \{\|\boldsymbol{\beta}-\boldsymbol{\beta}_0 \|<\epsilon_1\},\  E\left[\frac{f_{\boldsymbol{\beta}}}{f_{\boldsymbol{\beta_0}}}\right] < 1+ 2 E\left[\log\frac{f_{\boldsymbol{\beta}}}{f_{\boldsymbol{\beta_0}}}\right]<e^{- 2 E\left[\log\frac{f_{\boldsymbol{\beta}_0}}{f_{\boldsymbol{\beta}}}\right]}.\]
We can write $\{\Delta_n<\|\boldsymbol{\beta}- \boldsymbol{\beta}_0\|\leq \epsilon_1\} \subseteq \cup_{j\geq 1} A_{jn} $, where $A_{jn}=\{\boldsymbol{\beta}:  j\Delta_n<\|\boldsymbol{\beta}- \boldsymbol{\beta}_0\|\leq (j+1)\Delta_n \}$.  We note using Lemma 2(c) of \cite{sriram.rvr.ghosh.2013} and Assumption 3b that for sufficiently large $n$, for some constant $C_4>0$,
\[\sum_{i=1}^{n} E\left[\log\frac{f_{i\boldsymbol{\beta}_0}}{f_{i\boldsymbol{\beta}}}\right]\geq C_4\frac{nj^2\Delta^2_n}{4}, \ \forall \ \boldsymbol{\beta}\in A_{jn}\]
It follows that 
\begin{eqnarray*}
&&E\left[\int_{\Delta_n<\|\boldsymbol{\beta}- \boldsymbol{\beta}_0\|\leq \epsilon_1}(\sqrt{n}\|\boldsymbol{\beta}-\boldsymbol{\beta}_0 \|)^k  \prod_{i=1}^{n}\frac{f_{\boldsymbol{i\beta}}}{f_{\boldsymbol{i\beta_0}}}\pi(\boldsymbol{\beta})d\boldsymbol{\beta}\right]\\
&&\leq \sum_{j\geq 1}\int_{A_{jn}} (\sqrt{n}\|\boldsymbol{\beta}-\boldsymbol{\beta}_0 \|)^k  e^{- 2\sum_{i=1}^{n} E\left[\log\frac{f_{i\boldsymbol{\beta}_0}}{f_{i\boldsymbol{\beta}}}\right]}\pi(\boldsymbol{\beta})d\boldsymbol{\beta}\\
&&\leq (n\Delta^2_n)^{\frac{k}{2}} \sum_{j\geq 1} (j+1)^{\frac{k}{2}} e^{-C_4 \frac{nj^2\Delta^2_n}{4}}\Pi(A_{jn})\\
&&\leq C_5 \cdot(n\Delta^2_n)^{\frac{k}{2}} e^{-C_4 \frac{n\Delta^2_n}{4}} \cdot \Delta^p_n \sum_{j\geq 1} (j+1)^{\frac{k}{2}+d} e^{-C_4 \frac{n(j^2-1)\Delta^2_n}{4}}.
\end{eqnarray*}
The last expression uses the fact that  $\boldsymbol{\beta}\in \Re^p$ and hence $\Pi(A_{jn})$ is less than a constant multiple (say $C_5$) of $((j+1)\Delta_{n}^p$.
It is easy to see that the summation in the above expression is bounded for all $n$. Hence, for some constant $C_6$ and sufficiently large $n$, we will have
\begin{eqnarray}
\label{epsdelta_nr}
E\left[\int_{\Delta_n<\|\boldsymbol{\beta}- \boldsymbol{\beta}_0\|\leq \epsilon_1}(\sqrt{n}\|\boldsymbol{\beta}-\boldsymbol{\beta}_0 \|)^k  \prod_{i=1}^{n}\frac{f_{\boldsymbol{i\beta}}}{f_{\boldsymbol{i\beta_0}}}\pi(\boldsymbol{\beta})d\boldsymbol{\beta}\right]
\leq C_6\cdot(n\Delta^2_n)^{\frac{k}{2}} e^{-C_4 \frac{n\Delta^2_n}{4}} \cdot \Delta^p_n.
\end{eqnarray}
For some $C'_4>0$ (to be chosen later) let $B_n:=\{\boldsymbol{\beta}: \ \| \boldsymbol{\beta}-\boldsymbol{\beta_0}\|<C'_4 \Delta^2_n \}$. Then, using a similar argument as in the proof of Lemma 3.2 of \cite{Kleijn_van2012}, we get $\Pi(B_n)\geq K  \Delta_{n}^p$, for some constant K. Hence,
\begin{eqnarray}
\label{epsdelta_dr}
\int_{\Re^p}\prod_{i=1}^{n}\frac{f_{\boldsymbol{i\beta}}}{f_{\boldsymbol{i\beta_0}}}\pi(\boldsymbol{\beta})d\boldsymbol{\beta}\geq \int_{B_n}\prod_{i=1}^{n}\frac{f_{\boldsymbol{i\beta}}}{f_{\boldsymbol{i\beta_0}}}\pi(\boldsymbol{\beta})d\boldsymbol{\beta}\geq K  \Delta_{n}^p e^{-nC_4' \Delta^2_{n}}.
\end{eqnarray}
Choosing $C_4'=C_4/2$, equations (\ref{epsdelta_nr}) and (\ref{epsdelta_dr}) together imply
\begin{eqnarray}
\label{ExpDeltaneps}
E\left[\int_{\Delta_n<\|\boldsymbol{\beta}- \boldsymbol{\beta}_0\|\leq \epsilon_1}(\sqrt{n}\|\boldsymbol{\beta}-\boldsymbol{\beta}_0 \|)^k  d\Pi_n(\boldsymbol{\beta})\right] \leq C_6/K \cdot(n\Delta^2_n)^{\frac{k}{2}} e^{-C_4 \frac{n\Delta^2_n}{8}}\rightarrow \ 0 \ ( as \ n\rightarrow \infty).
\end{eqnarray}
Equations (\ref{ExpGcomp}), (\ref{ExpGeps})and (\ref{ExpDeltaneps}) together give the result
\[E\left[\int_{\|\boldsymbol{\beta}- \boldsymbol{\beta}_0\|>\Delta_n}(\sqrt{n}\|\boldsymbol{\beta}-\boldsymbol{\beta}_0 \|)^k  d\Pi_n(\boldsymbol{\beta})\right] \rightarrow \ 0.\]
\end{proof}

\begin{proof}[{\bf \underline{Proof of Lemma 2}}]
Suppose for a vector ${\bf L}$, let  $\theta:={\bf L}^{T}\boldsymbol{\beta}$,  $\theta_0:={\bf L}^{T}\boldsymbol{\beta}_0$, $\theta^{M}_n:={\bf L}^{T}\boldsymbol{\beta}^{M}_n$ and $\tilde{\theta}_n:={\bf L}^{T}\tilde{\boldsymbol{\beta}}_n$. By abuse of notation, we will continue to use $\Pi_n()$ to denote the posterior distribution of $\theta$ from Step 1.
Since $\theta$ is just a linear combination of $\boldsymbol{\beta}$, Lemma 1 implies (for $k\in\{0,1,2\}$)
\begin{equation}
\label{postconvsqL}
 \int_{\{ \theta : \sqrt{n}|\theta-\theta_0|>\frac{M_n}{2} \}} (\sqrt{n}|\theta-\theta_0|)^k d\Pi_n(\theta)  \rightarrow \ 0 \mbox{ in  probability } [P].
\end{equation}
Note that 
\begin{eqnarray}
 &&\int_{\{ \theta : \sqrt{n}|\theta-\theta_0|>\frac{M_n}{2} \}}n|\theta-\theta^M_n|^2 d\Pi_n(\theta) \nonumber\\
 &&\leq 2\int_{\{ \theta : \sqrt{n}|\theta-\theta_0|>\frac{M_n}{2} \}} n|\theta-\theta_0|^2 d\Pi_n(\theta)+ 2n |\theta^M_n-\theta_0|^2 \cdot\Pi_n(\sqrt{n}|\theta-\theta_0|>\frac{M_n}{2})\label{eqn20}
\end{eqnarray}
Using equation  (\ref{postconvsqL}) and the fact $n|\theta^M_n-\theta_0|^2$ is bounded in probability, we get that the right hand side of equation  (\ref{eqn20}) converges to zero in probability. Hence, 
\begin{eqnarray}
\label{postconvsqL2}
 &&\int_{\{ \theta : \sqrt{n}|\theta-\theta_0|>\frac{M_n}{2} \}}n|\theta-\theta^M_n|^2 d\Pi_n(\theta) \rightarrow \ 0 \mbox{ in  probability } [P]
\end{eqnarray}
Further,
\begin{eqnarray}
\label{postconvsqL3}
&\int_{\{ \theta : \sqrt{n}|\theta-\theta^M_n|>M_n \}}n|\theta-\theta^M_n|^2 d\Pi_n(\theta) \leq  \int_{\{ \theta : \sqrt{n}|\theta-\theta_0|>\frac{M_n}{2} \}} n|\theta-\theta^M_n|^2 d\Pi_n(\theta) \nonumber \\
& + \int_{\{ \theta : \sqrt{n}|\theta-\theta_0|\leq \frac{M_n}{2}, \ \sqrt{n}|\theta-\theta^M_n|> M_n  \}}n|\theta-\theta^M_n|^2 d\Pi_n(\theta)  
\end{eqnarray}
First term on right hand side of (\ref{postconvsqL3}) is same as that in equation (\ref{postconvsqL2}). Since $\sqrt{n}|\theta-\theta_0|\leq \frac{M_n}{2}$ and $ \sqrt{n}|\theta-\theta^M_n|> M_n $ would imply $\sqrt{n}|\theta^M_n-\theta_0|> \frac{M_n}{2}$, for any $\epsilon>0$, for the second term, we can write ,
\[P\left(\int_{\{ \theta : \sqrt{n}|\theta-\theta_0|\leq \frac{M_n}{2}, \ \sqrt{n}|\theta-\theta^M_n|> M_n  \}}n|\theta-\theta^M_n|^2 d\Pi_n(\theta) >\epsilon\right) \leq P(\sqrt{n}|\theta^M_n-\theta_0|> \frac{M_n}{2})\]
Since the right hand side goes to zero as $n\rightarrow \infty$, the second term of equation (\ref{postconvsqL3}) converges to zero in probability. Therefore, we have established that
\begin{eqnarray}
\label{postconvsqL4}
&\int_{\{ \theta : \sqrt{n}|\theta-\theta^M_n|>M_n \}}n|\theta-\theta^M_n|^2 d\Pi_n(\theta) \rightarrow \ 0 \ \mbox{ in probability } [P]
\end{eqnarray}
Now, let $Q_n$ denote the distribution of $\sqrt{n}(\theta-\theta^M_n)$ under $\theta\sim \Pi_n$.  Let $Q$ denote $N(0, {\bf L}^TV^{-1}{\bf L})$. Then Theorem 1 implies that $Q_n$ converges to $Q$ with respect to the total variation norm. Therefore,
\[\mbox{ for any fixed } M, \int_{|x|\leq M}x^2 dQ_n \rightarrow \int_{|x|\leq M}x^2 dQ \mbox{ as } n\rightarrow \infty \mbox{, and } \int_{|x|\leq M}x^2 dQ \rightarrow \int x^2 dQ \mbox{ as } M\rightarrow \infty.\]
So, there exists a sequence $\{M_n\}$ such that $\int_{|x|\leq M_n}x^2 dQ_n \rightarrow \int x^2 dQ \mbox{ as } n\rightarrow \infty$. For such a sequence, using equation \ref{postconvsqL4}, we get $\int x^2 dQ_n \rightarrow \int x^2 dQ \mbox{ as } n\rightarrow \infty.$ Equivalently,
\begin{equation}
\label{convmoment}
\int n(\theta-\theta^M_n)^2 d\Pi_n \rightarrow  {\bf L} V^{-1}{\bf L} \mbox{ as } n\rightarrow \infty.
\end{equation}
Following the same approach as above, we can also conclude that:
\begin{eqnarray}
\int |n(\theta-\theta^M_n)| d\Pi_n \rightarrow  \int |x|dQ(x)\mbox{ as } n\rightarrow \infty. \label{convabs}\\
\int (n(\theta-\theta^M_n))^{+} d\Pi_n \rightarrow  \int x^{+}dQ(x)\mbox{ as } n\rightarrow \infty. \label{convplus}
\end{eqnarray}
It is now easy to see that equations (\ref{convabs}) and (\ref{convplus}) imply that
\[\sqrt{n}({\bf L}^{T}\boldsymbol{\beta}^M_n - {\bf L}^{T}\boldsymbol{\tilde{\beta}}_n)\rightarrow \ 0 \ \mbox{ in probability } [P]\]
This along with equation (\ref{convmoment}) implies that
\[\int n(\theta-\tilde{\theta}_n)^2 d\Pi_n \rightarrow  {\bf L} V^{-1}{\bf L} \mbox{ as } n\rightarrow \infty.\]
Since the above two results hold for any vector ${\bf L}$, statements (a) and (b) of Lemma 2 follow.
\end{proof}

\newpage
\begin{proof}[{\bf \underline{Proof of Lemma 3}}]
Recall $Z_i=Y_i-{\bf X}^{T}_i\boldsymbol{\beta}_0$. Let $P_i(\cdot)$ denote the cumulative distribution function for the p.d.f $p_i(\cdot)$. In the lines of proof of Theorem 4.1 in \cite{Koenker2005}, we write
\begin{eqnarray*}
&&U_{n}(\boldsymbol{\delta}):= \log \frac{f^{(n)}_{\boldsymbol{\beta}_0+\frac{\boldsymbol{\delta}}{\sqrt{n}}}}{f^{(n)}_{\boldsymbol{\beta}_0}} = U_{1n}(\boldsymbol{\delta}) + E[U_{2n}] + (U_{2n}(\boldsymbol{\delta})-E[U_{2n}(\boldsymbol{\delta})]) \\
\mbox{ where}, && U_{1n}(\boldsymbol{\delta}):=-\boldsymbol{\delta}^{T}\frac{1}{\sqrt{n}}\sum_{i=1}^{n}(\tau-I_{\{Z_i\leq 0\}}){\bf X}_i, \mbox{ and } \\
&& U_{2n}(\boldsymbol{\delta}):=\sum_{i=1}^{n}\int_{0}^{\frac{{\bf  X}^{T}_i\boldsymbol{\delta}}{\sqrt{n}}}(I_{\{ Z_i\leq s\}}-I_{\{Z_i\leq 0\}})ds=\sum_{i=1}^{n}U_{2ni}(\boldsymbol{\delta}).
\end{eqnarray*}
Further, by Taylor's formula, 
\begin{eqnarray*}
 E(U_{2n}(\boldsymbol{\delta}))&& = \sum_{i=1}^{n}\int_{0}^{\frac{{\bf  X}^{T}_i\boldsymbol{\delta}}{\sqrt{n}}}(P_i({\bf  X}^{T}_i\boldsymbol{\beta}_0+s)-P_i({\bf  X}^{T}_i\boldsymbol{\beta}_0))ds= \sum_{i=1}^{n}\int_{0}^{\frac{{\bf  X}^{T}_i\boldsymbol{\delta}}{\sqrt{n}}}p_i({\bf  X}^{T}_i\boldsymbol{\beta}_0+r_{in}(s))\cdot s \ ds,
\end{eqnarray*}
where $0\leq r_{in}(s)\leq s \leq \left\vert \frac{{\bf X}^{T}_i\boldsymbol{\delta}}{\sqrt{n}}\right\vert$. Therefore, by Assumptions {\bf 2} and {\bf 5}, we have
\begin{eqnarray*}
 &&\left\vert E(U_{2n}(\boldsymbol{\delta})) - \frac{1}{2n}\sum_{i=1}^{n}p_{i}({\bf X}^{T}_i\boldsymbol{\beta}_0)\boldsymbol{\delta}^{T}{\bf X}_i{\bf X}^{T}_i\boldsymbol{\delta}\right\vert\\
  && \leq \sum_{i=1}^{n}\int_{0}^{\frac{{\bf  X}^{T}_i\boldsymbol{\delta}}{\sqrt{n}}} C \left\vert r_{in}(s) \right\vert^{\eta}\cdot s \ ds \leq C \left(\max_{i\geq 1} \frac{|{\bf X}^{T}_i \boldsymbol{\delta}|}{\sqrt{n}}\right)^{\eta} \cdot \boldsymbol{\delta}^{T}\frac{1}{n}\sum_{i=1}^{n}{\bf X}_i{\bf X}^{T}_i\boldsymbol{\delta}.
\end{eqnarray*}
It follows that,
\begin{eqnarray}
\sup_{\boldsymbol{\delta}\in K}\left\vert E(U_{2n}(\boldsymbol{\delta})) - \frac{1}{2n}\sum_{i=1}^{n}p_{i}({\bf X}^{T}_i\boldsymbol{\beta}_0)\boldsymbol{\delta}^{T}{\bf X}_i{\bf X}^{T}_i\boldsymbol{\delta}\right\vert \rightarrow \ 0 \ \mbox{ in probability  } [P].
\end{eqnarray}
To complete the proof, it is enough to show  $\sup_{\boldsymbol{\delta }\in K } \left\vert U_{2n}(\boldsymbol{\delta})- E\left[U_{2n}(\boldsymbol{\delta})\right]\right\vert$  $\rightarrow \ 0 $ in probability $[P]$. To show this, let $\boldsymbol{\delta}_i=arg\max_{\boldsymbol{\delta}\in K} \left\vert U_{2ni}(\boldsymbol{\delta})- E\left[U_{2ni}(\boldsymbol{\delta})\right]\right\vert$.  $\boldsymbol{\delta}_i$ is possibly random since it can depend on $Z_i$, but is well defined since $U_{2ni}(\boldsymbol{\delta})$ is a continuous (random) function on a compact set $K$. Then,
\begin{eqnarray}
&&P\left(\sup_{\boldsymbol{\delta }\in K } \left\vert U_{2n}(\boldsymbol{\delta})- E\left[U_{2n}(\boldsymbol{\delta})\right]\right\vert>\epsilon \right) \leq  P\left(\sum_{i=1}^{n}\sup_{\boldsymbol{\delta }\in K } \left\vert U_{2ni}(\boldsymbol{\delta})- E\left[U_{2ni}(\boldsymbol{\delta})\right]\right\vert>\epsilon \right) \nonumber\\
&&=  P\left(\sum_{i=1}^{n} \left\vert U_{2ni}(\boldsymbol{\delta}_i)- E\left[U_{2ni}(\boldsymbol{\delta}_i)\right]\right\vert>\epsilon \right)\leq \sum_{i=1}^{n}\frac{E\left( U^2_{2ni}(\boldsymbol{\delta}_i) \right)}{\epsilon^2}. \label{previneq} 
\end{eqnarray}
The last step uses Chebyhev's inequality. Further, let $G= \sup_{\delta\in K} \sup_{i\geq 1}  \left\vert {\bf  X}^{T}_i\boldsymbol{\delta} \right\vert$. In particular,  $\left\vert {\bf  X}^{T}_i\boldsymbol{\delta_i}\right\vert\leq G $.  
Without loss of generality, if ${\bf X}^{T}_i\boldsymbol{\delta} >0 , \ \forall \ i$,  the right hand side of Equation (\ref{previneq}) is
\begin{eqnarray*}
&&\leq \sum_{i=1}^{n}\frac{E\left(\int_{0}^{\frac{{\bf  X}^{T}_i\boldsymbol{\delta}_i}{\sqrt{n}}}(I_{\{ Z_i\leq s\}}-I_{\{Z_i\leq 0\}})ds\right)^2}{\epsilon^2}\leq \sum_{i=1}^{n}\frac{E\left(\int_{0}^{\frac{G}{\sqrt{n}}}(I_{\{ Z_i\leq s\}}-I_{\{Z_i\leq 0\}})ds\right)^2}{\epsilon^2}\\
\end{eqnarray*}
\begin{eqnarray*}
&&\leq \sum_{i=1}^{n}\frac{G^2}{n}\frac{E\left(I_{\{ Z_i\leq \frac{G}{\sqrt{n}}\}}-I_{\{Z_i\leq 0\}}\right)^2ds}{\epsilon^2}\leq \frac{G^2}{n\epsilon^2}\sum_{i=1}^{n}\left(P_i({\bf X}^{T}_i\boldsymbol{\beta}_0+ \frac{G}{\sqrt{n}})-P_i({\bf X}^{T}_i\boldsymbol{\beta}_0)\right) \\
&& \leq \frac{G^2}{n\epsilon^2} \sum_{i=1}^{n}\left\vert p_i\left({\bf X}^{T}_i \boldsymbol{\beta}_0+ r_{in}\right)\right\vert\frac{G}{\sqrt{n}}, \mbox{ for some }|r_{in}| \leq \frac{G}{\sqrt{n}} \mbox{ using Taylor's formula}. 
\end{eqnarray*}
 A consequence of assumptions {\bf 2} and {\bf 5} is that $\{p_i({\bf X}^{T}_i\boldsymbol{\beta}_0 +r_{in}), \ i\geq 1\}$ are uniformly bounded. Hence,  we can conclude that $P\left(\sup_{\boldsymbol{\delta }\in K } \left\vert U_{2n}(\boldsymbol{\delta})- E\left[U_{2n}(\boldsymbol{\delta})\right]\right\vert>\epsilon \right)\rightarrow 0$.
\end{proof}

{\small
\begin{singlespace}
\bibliographystyle{asa}

\bibliography{ref_MisspecALD}
\end{singlespace}

\end{document}